\documentclass[lettersize,journal]{IEEEtran}
\usepackage{amsfonts}
\usepackage{times}
\usepackage{graphicx}
\usepackage{latexsym}
\usepackage{dsfont}
\usepackage{amssymb}
\usepackage{amsmath}
\usepackage{cite}
\usepackage{verbatim}
\usepackage{subfigure}

\newcommand{\figref}[1]{{Fig.}~\ref{#1}}

\def\bb0{{\mathbb{0}}}

\def\bb{{\mathbf{b}}}

\def\b0{{\mathbf{0}}}

\def\sf0{{\mathsf{0}}}

\usepackage{fancyhdr}

\fancypagestyle{firstpage}{
	\fancyhf{}
	\fancyhead[HL]{\fontsize{9pt}{11pt}\selectfont 2023 IEEE. Personal use of this material is permitted. Permission from IEEE must be obtained for all other uses, in any current or future media, including reprinting/republishing this material for advertising or promotional purposes, creating new collective works, for resale or redistribution to servers or lists, or reuse of any copyrighted component of this work in other works.}
	
}

\usepackage{epstopdf}
\usepackage{enumerate}
\usepackage{algorithmicx}
\usepackage{algorithm}
\usepackage{amsmath}
\usepackage[noend]{algpseudocode}
\usepackage{float}
\usepackage{hyperref}
\usepackage{color}
\usepackage{makeidx}
\usepackage{bbm}
\usepackage{graphicx}
\usepackage{lipsum}
\usepackage{subfigure}
\usepackage{tablefootnote}
\usepackage{multirow}
\usepackage{multicol}
\usepackage{balance}
\usepackage{booktabs}
\usepackage{soul}
\usepackage[normalem]{ulem}
\usepackage{xcolor}
\usepackage{gensymb}

\newcommand{\sref}[1]{{Section}~\ref{#1}}

\newcommand{\comm}[1]{}

\begin{document}
    
	\title{DeepSense 6G: A Large-Scale Real-World Multi-Modal Sensing and Communication Dataset}
	
	\author{Ahmed Alkhateeb, Gouranga Charan, Tawfik Osman, Andrew Hredzak, João Morais, \\ Umut Demirhan, and Nikhil Srinivas  \thanks{The authors are with the School of Electrical, Computer, and Energy Engineering at Arizona State University. Emails: \{alkhateeb, gcharan, tmosman, ahredzak, joao, udemirhan, tvsrini1\}@asu.edu.}} 
	
	\maketitle
	\thispagestyle{firstpage}


	\begin{abstract}
		This article presents the DeepSense 6G dataset, which is a large-scale dataset based on real-world measurements of co-existing multi-modal sensing and communication data. The DeepSense 6G dataset is built to advance deep learning research in a wide range of applications in the intersection of multi-modal sensing, communication, and positioning. This article provides a detailed overview of the DeepSense dataset structure, adopted testbeds, data collection and processing methodology, deployment scenarios, and example applications, with the objective of facilitating the adoption and reproducibility of multi-modal sensing and communication datasets. 
	\end{abstract}
	
	\section{Introduction} \label{sec:intro}
	The synergy between communication, multi-modal sensing, and positioning is envisioned as a defining characteristic of future wireless systems in 6G and beyond \cite{Zhang_radar_21,9737357,9705498, Charan21_TVT, Zecchin, Charan_ML22, 9771564,9939167}. These systems will likely either implement co-existing communication and sensing functions or utilize one to aid the other. This is particularly motivated by the move to higher frequency bands where the wide bandwidth and large antenna arrays are attractive features for both communication and sensing. This synergy has recently been the driver for key research directions such as multi-modal sensing-aided communication \cite{Charan21_TVT, Zecchin, Charan_ML22, 9771564,9939167,Ali_2020}, integrated sensing and communication \cite{Zhang_radar_21,9771564,9705498}, and communication-aided positioning \cite{Kim_20}. Machine learning and deep learning have the potential to play fundamental roles in many of these problems \cite{Demirhan_mgazine_radar, Charan21_TVT, Charan_ML22, Zecchin}. The ability, however, to develop and evaluate deep learning solutions depends on the availability of large enough datasets.  
	
	Multiple synthetic datasets have recently been developed with the objective of facilitating machine learning research in wireless communications. For example, in \cite{Alkhateeb2019}, the authors developed the DeepMIMO dataset, which is a parametric data generation framework based on ray-tracing simulations targeted mainly for MIMO research. In \cite{Alrabeiah2020b}, the ViWi dataset was developed to generate both wireless and visual/LiDAR data. While providing a path for initial algorithm development, synthetic datasets may not be sufficient for obtaining realistic insights about the performance of the machine learning algorithms in real-world deployments. 
	
	With this motivation, we developed the DeepSense 6G dataset \footnote{The DeepSense 6G dataset is publicly available at \url{https://deepsense6g.net/}. Some of the DeepSense 6G scenarios (datasets) have been used before to enable the deep learning research in \cite{Charan_ML22,Demirhan_mgazine_radar,Charan2022}.}, the world's first large-scale real-world multi-modal sensing and communication dataset. The DeepSense 6G dataset is (i) a large-scale dataset of \textbf{more than 1 million data points}, (ii) based on \textbf{real-world measurements}. The dataset comprises co-existing and synchronized \textbf{multi-modal sensing and communication data} and is organized in a \textbf{collection of 40+ scenarios} that cover \textbf{diverse} deployment use cases such as vehicle to infrastructure, vehicle to vehicle, pedestrians, drone communication, fixed-wireless, and indoor use cases. With that, the DeepSense 6G dataset has the potential to enable a wide range of applications in the intersection of communication, sensing, and positioning. 
	
	The goal of this article is to provide an overview of the DeepSense dataset structure, testbeds, data collection methodology, scenarios, and enabled applications. In particular, the article describes DeepSense in detail to facilitate the adoption and reproducibility of the DeepSense testbeds, data collection process, and datasets.  
	
	\begin{figure*}[!t]
		\centering
		\includegraphics[width=1\linewidth]{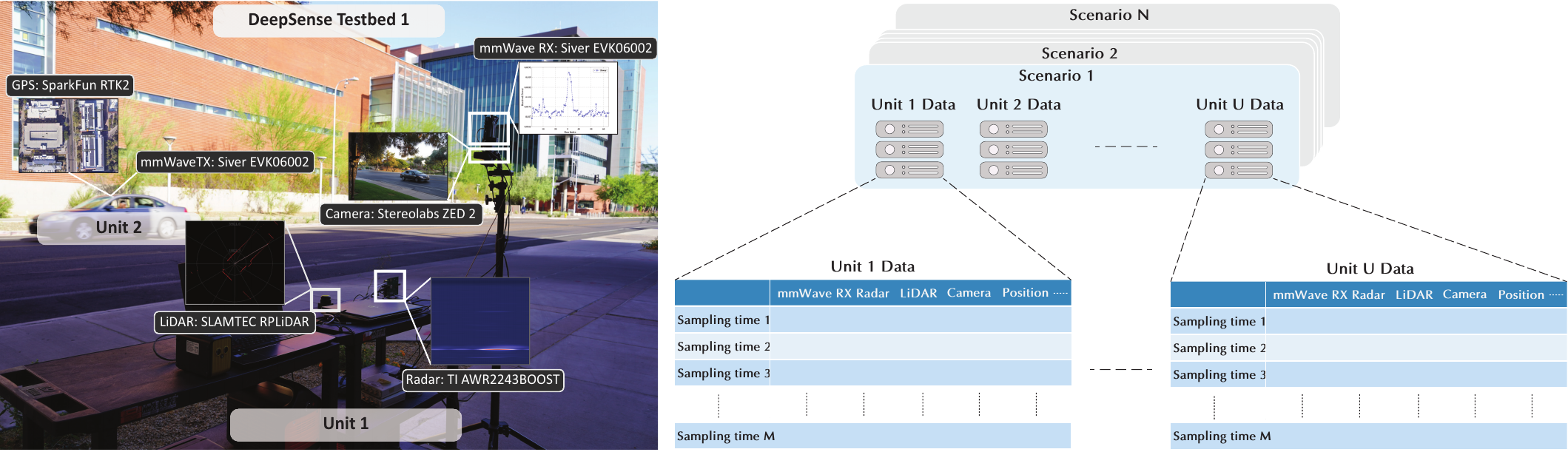}
		\caption{ This figure presents the DeepSense 6G testbed 1 and the different sensing modalities. It consists of two units: Unit 1 (a stationary unit), which acts as the basestation, and unit 2 (a vehicle), which represents the mobile user. It also shows the final scenario structure, where the data is eventually stored as a sequence of data groups with each group containing data collected from all the sensors at the same sampling interval.  }
		\label{fig:data_structure}
	\end{figure*}
	
	\section{Why DeepSense 6G?} \label{sec:motivation}
	As discussed in the introduction, machine and deep learning are rapidly finding applications in the communication, sensing, and positioning systems and their intersection. The ability, however, to develop and adequately evaluate the performance of machine/deep learning approaches and derive meaningful conclusions is conditioned on having a dataset with the following key characteristics: 
	\begin{itemize}
		\item \textbf{Co-existing sensing and communication:} To enable the targeted research directions, this dataset should comprise \textit{co-existing} sensing and communication data that are synchronized to a sufficient extent. For example, both the wireless communication and sensing measurements need to be collected within the same coherence time. 
		\item \textbf{Multi-modal sensing data:} By sensing, we do not mean only radar, but more broadly multi-modal sensing. This includes, for example, visual data, LiDAR sensory data, GPS positions, and weather data captured by sensors deployed at the infrastructure or mobile devices. Recent research has shown that a very interesting synergy exists between multi-modal sensing and communication \cite{Demirhan_mgazine_radar,Charan21_TVT,Charan_ML22,Zecchin}, which can be leveraged to benefit both of them. 
		\item \textbf{Real-world measurements:} While using synthetic data, e.g. \cite{Alkhateeb2019,Alrabeiah2020b}, can provide initial insights, advancing the machine learning research and development towards real-world adoption and deployment requires the availability of real-world datasets. This is essential as some real-world and practical imperfections are very hard to model and capture in the synthetic datasets. 
		\item \textbf{Large-scale:} Developing deep learning solutions that are scalable and robust to data distribution shifts (due to changes in the environment or deployment) requires the availability of a large-scale dataset. Therefore, the ideal dataset should have a large number of data points and should have sufficient variance in both the communication and the multi-modal sensing measurements. 
		\item \textbf{Scalable to various scenarios:} The applications and use cases of integrated multi-modal sensing and communication are countless. From the deployment scenarios (vehicle to infrastructure (V2I), vehicle to vehicle (V2V), indoor or outdoor, etc.) to the adopted devices (sub-6GHz massive MIMO, mmWave, reconfigurable intelligent surfaces, etc.), it is not feasible to have one data collection that covers all applications and use cases. Therefore, the dataset definition and structure have to be scalable to enable the continuous growth of the dataset. 
	\end{itemize}
	
	To achieve these objectives, we built the DeepSense 6G dataset, which is a large-scale real-world dataset comprising co-existing multi-modal sensing and communication data covering various use cases and enabling many applications in the interplay of communications, sensing, and positioning.

	\section{DeepSense Dataset Structure} \label{sec:structure}
	The DeepSense 6G dataset is \textit{a collection of scenarios}. Each scenario is a standalone dataset comprising the multi-modal sensing and communication data collected in one, typically long, data collection session and formatted in a generic way that is unified across all the scenarios. The data collection of each scenario is planned to cover an important deployment scenario and enable one or more applications (the available scenarios will be discussed in \sref{sec:deepsense_scenarios}). This simple structure achieves a few important objectives: (i) It enables the scalability of the DeepSense datasets as it can keep growing to address more use cases by adding more scenarios. (ii) The unified way is offering the data in each scenario facilitates combining this data from multiple scenarios (e.g., multiple sites, different times of the day, etc.) to build a bigger dataset with the targeted diversity. (iii) It simplifies the access, the definition, and the reproducibility of the dataset, which are critical goals for a useful deep learning dataset.

	\textbf{Data structure in each scenario:} Each DeepSense 6G scenario has the multi-modal sensing and communication data collected in one data collection session. As will be described in \sref{sec:testbeds}, each scenario uses a testbed that consists of a number of units (for example, a basestation and a mobile user unit in a V2I scenario). Each unit is equipped with a number of sensors such as wireless receivers, radar, LiDAR, cameras, and GPS receivers. The data collection and processing are designed such that the data from all the sensors and units are collected and synchronized to the same sampling interval. Therefore, as shown in \figref{fig:data_structure}, the scenario data is eventually structured as a sequence of data groups, where each data group has the sensory data collected from all the sensors at the same sampling interval. To facilitate accessing the data, the DeepSense 6G website dedicates a page for each scenario with a detailed description of the adopted testbed, deployment, available modalities, and scenario folder structure.

	\begin{figure*}[!t]
		\centering
		\includegraphics[width=1.0\linewidth]{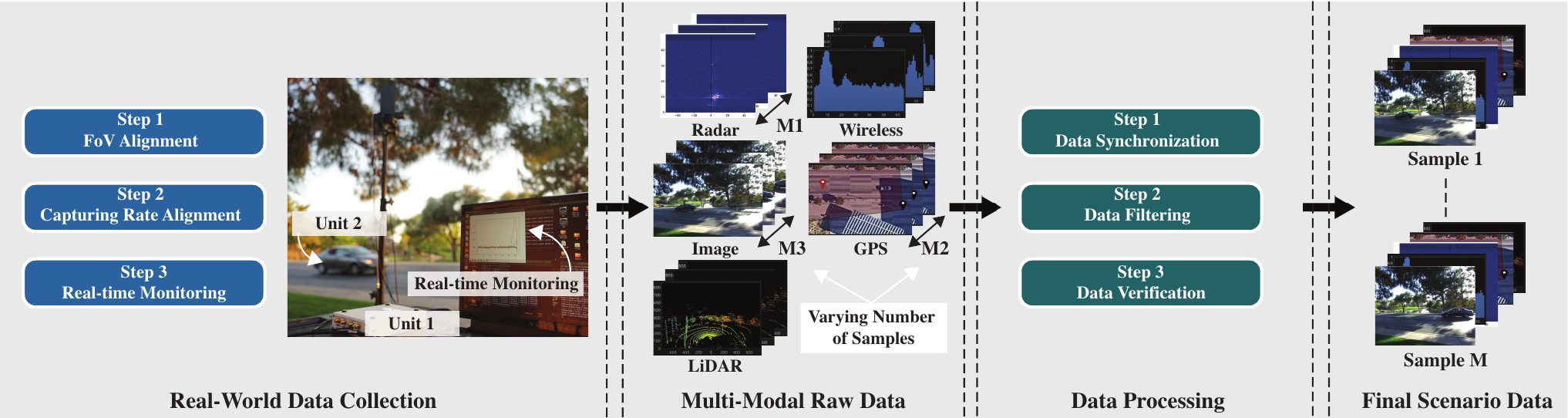}
		\caption{ The DeepSense 6G dataset is a collection of scenarios, where each scenario consists of the multi-modal sensing and communication data collected in one long data collection session. This figure outlines the different steps required to generate the final development (scenario) dataset after each data collection session. The process comprises three steps: (i) Real-world data collection, (ii) data processing, and (iii) data filtering. }
		\label{fig:scenario_data_gen}
	\end{figure*}

	\section{DeepSense Testbeds} \label{sec:testbeds}
	
	To enable the scalability of the dataset and to cover various scenarios, we adopt a series of testbeds in the collection of the DeepSense scenarios. More specifically, each DeepSense testbed is used to collect the data of one or more scenarios. Further, we adopt a unified and modular approach in building and describing these testbeds. Each testbed is composed of a number of units; each unit is a self-sustained module equipped with: (i) A set of wireless and environmental sensors such as wireless communication transceiver in one or more bands (such as mmWave phased arrays or sub-6GHz MIMO). The units are also equipped with a suite of sensors, such as an RGB camera, $2$D/$3$D LiDAR, radar, and GPS RTK kit, to capture additional information about the surrounding environment. (ii) A processing unit to initialize and configure the sensors as well as store the collected sensing data. For each DeepSense testbed, the units are equipped with the sensors that suit the targeted application(s). Table \ref{tab:data_statistics} provides a summary of the testbeds and their adopted sensors.

	\textbf{Example - DeepSense Testbed 5:} Given the modular design of the testbeds, providing one example may help clarify the testbed capabilities and enable the reproduction of the testbeds and datasets. For this objective, we now provide more details about a representative testbed, namely Testbed 5. This testbed targets vehicle-to-infrastructure (V2I) applications. Therefore, we designed this testbed to have two units: Unit 1 (a stationary unit), which acts as the basestation, and unit 2 (a vehicle), which represents the mobile user. Unit 1 is equipped with the following devices: 
	\begin{itemize}
		\item \textbf{A mmWave receiver} that includes a 60GHz RF front-end with a 16-element uniform linear (phased) array (from SIVERS semiconductors). The mmWave receiver adopts a combining beamforming codebook of 64 beams that uniformly scan 90$\degree$ field-of-view. 
		\item \textbf{RGB Camera}  with 110$\degree$ field-of-view and 30 frames-per-second (ZED2 from StereoLabs).  
		\item \textbf{3D LiDAR} with 32 vertical and 1024 horizontal channels and with 120 m range and 20 Hz frame rate (from Ouster).
		\item \textbf{Radar} that adopts frequency modulated continuous wave (FMCW). It operates at 76-81GHz and has a maximum range of 100 m (from Texas Instruments). 
		\item \textbf{GPS receiver} that uses real-time kinematic positioning (RTK) technology and a 10 Hz rate (from SparkFun).
	\end{itemize} 
	
	The mobile user (unit 2) employs a mmWave transmitter with a $60$GHz quasi-omni antenna (from SIVERS semiconductors) and the same GPS device used by the base station. The data collected at each time instant comprises the GPS position of the user, the RGB image, radar I/Q samples, LiDAR point cloud, and the $64$-element power vectors that correspond to the mmWave beam training done at the basestation.

	\section{Data Collection and Processing Framework} \label{sec:data_collection}
	As described in Section~\ref{sec:structure}, the DeepSense 6G dataset is a collection of scenarios, where each scenario consists of the multi-modal sensing and communication data collected in one long data collection session. Building the dataset of a new DeepSense 6G scenario comprises two main phases, i.e., data collection and data processing. Next, we present an overview of the two phases. Fig.~\ref{fig:scenario_data_gen} illustrates the different steps involved in generating the final development dataset. 
	
	\begin{figure*}[!t]
		\centering
		\includegraphics[width=1.0\linewidth]{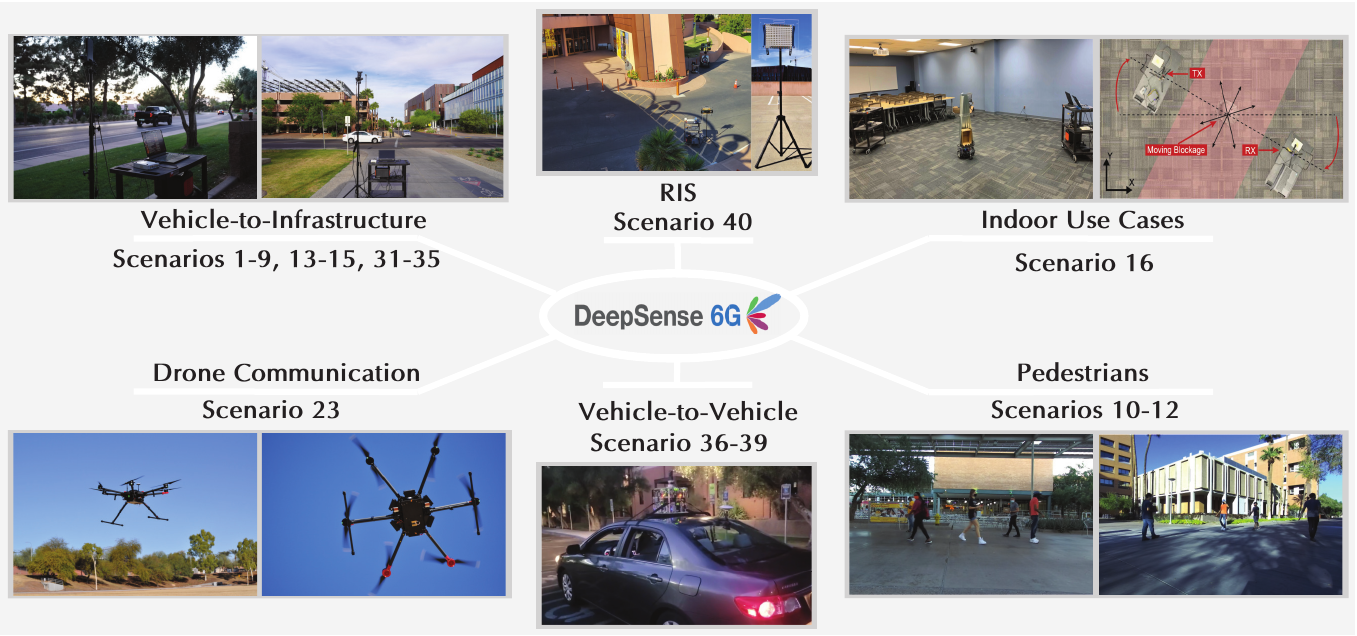}
		\caption{ The DeepSense 6G dataset comprises $40$+ scenarios with realistic datasets to encourage the development of novel applications. The dataset is collected to enable various deployment scenarios. This figure presents the different deployment scenarios included in the DeepSense 6G dataset. As shown here, the dataset covers a wide range of deployments such as vehicle to infrastructure, vehicle to vehicle, drone communication, reconfigurable intelligent surfaces, pedestrians, indoor use cases, to name a few. }
		\label{fig:scenarios}
	\end{figure*}

	\subsection{Data Collection} \label{subsec:data_collection} 
	
	The data collection process comprises two main phases: Off-field planning and on-field collection. Similar to any experimental design, the offline planning phase is extremely critical to the success of any data collection. It primarily involves clearly defining the set of objectives, such as the number of units needed and the different sensors that must be deployed in each unit. It also includes converging on the time and location of the data collection. The next step is the actual on-field data collection. It involves deploying the DeepSense 6G testbed in the pre-determined location. 
	
	\textbf{FoV alignment:} The different sensors, such as the mmWave phased array, RGB camera, LiDAR, and radar, have varying field-of-view (FoV) and range. For example: in the DeepSense testbed $5$, the RGB camera has an FoV $110{\degree}$, whereas the phased array has $90\degree$ FoV. In order to achieve a high-quality multi-modal dataset with multiple sensors, it is essential to align these different sensors before initiating the data collection process. For this, the testbed has provisions for some calibration steps to align the sensors considering the different ranges and FoVs. 
	
	\textbf{Capturing rate alignment:} The different sensing elements, such as the mmWave receiver, RGB camera, GPS RTK kit, LiDAR, and radar, have varying data capture rates. For instance, the mmWave receiver runs at $10$Hz, the camera can capture data at $30$Hz, whereas the 3D LiDAR can complete a maximum of $20$ sweeps in a second. Therefore, in order to achieve meaningful data, cross-modality data alignment is necessary. We undertake a series of actions in order to achieve the necessary alignment. For example, the exposure of the camera is triggered right after the beam sweeping is completed. Since the camera’s exposure time is nearly instantaneous, this method generally yields good data alignment between the camera and the mmWave receiver. Further, the UTC timestamp of the data capture is stored for all modalities. The timestamp is then utilized to align the different data modalities during the post-processing resulting in coherent data for each time step. 
	
	\textbf{Real-time monitoring:} The final step in the on-field data collection is the real-time monitoring of the collected data. Apart from the manual check, we have also implemented data plotting capabilities in the testbed to enable us to monitor the data in real-time to ensure their quality and correctness.

	\subsection{Data Processing} \label{subsec:data_processing}
	
	The initial data collected using the testbed is referred to as the raw data. The raw data further undergoes a post-processing pipeline to generate the final development dataset. The post-processing steps include data synchronization and filtering. Next, we present the details of each of the steps.

	\textbf{Data Synchronization:} The adopted sensing elements have different data capture rates, which might lead to an incoherent dataset. As described in Section~\ref{subsec:data_collection}, the data capture flow in the testbed is designed to limit such misalignment. However, the provision in the testbed alone might not be sufficient. To avoid such inconsistency in the final dataset, the testbed is designed to record the UTC timestamp along with the actual captured data. The saved timestamps act as the pivot point and help in aligning the different data modalities. We have developed scripts that can utilize the saved timestamps and generate a synchronized and aligned multi-modal dataset.

	\textbf{Data Filtering:} Depending upon the particular scenario, an optional data filtering step might be required. It is essential to highlight here that during the data collection process, once the collection has been started, the testbed continuously captures data until the process is stopped manually. Therefore, for any scenario with a mobile user, the continuous data collection process might result in samples where the user is outside the FoV of the basestation. For instance, in a V2I scenario developed to study sensing-aided beam prediction,  only the data samples with the mobile user (unit 2) in the FoV of the basestation are preferred. Therefore, such a filtering process is needed to ensure that the final development dataset has only the desired data samples.

	\textbf{Data Verification:} The final step of generating the development dataset is verifying the synchronized and filtered data. For that, we have developed in-house tools that assist in this verification process. One such tool is the data visualizer, which plots all the data modalities in the same GUI. It helps to filter out synchronization issues between different modalities and identify any missing or corrupted data. Overall, these tools help us ensure and maintain the highest quality of the datasets.

	\section{DeepSense 6G Scenarios} \label{sec:deepsense_scenarios}

	\begin{table*}[!t]
		\caption{DeepSense 6G Dataset: Testbeds, Scenarios, and Example Applications}
		\centering
		\setlength{\tabcolsep}{5pt}
		\renewcommand{\arraystretch}{1.5}
		\begin{tabular}{|c|c|c|c|c|c|}
			\hline
			\textbf{Testbed}   & \textbf{Scenarios} & \textbf{Deployment}             & \textbf{Data Modalities}                                                                                           & \textbf{No. of Samples} & \textbf{Example Applications}                                                                                   \\ \hline
			\multirow{4}{*}{1} & 1-7, 13-15         & \multirow{3}{*}{V2I}            & RGB images, GPS,  mmWave receive power                                                                             & 14,322            & \begin{tabular}[c]{@{}c@{}}Beam prediction, \\ user identification, positioning \end{tabular}              \\ \cline{2-2} \cline{4-6} 
			& 8                  &                                 & \begin{tabular}[c]{@{}c@{}}RGB images, GPS,  mmWave receive power,\\ and 2D LiDAR measurements\end{tabular}        & 4,043              & Beam prediction                                                                              \\ \cline{2-2} \cline{4-6} 
			& 9                  &                                 & \begin{tabular}[c]{@{}c@{}}RGB images, GPS,  mmWave receive power,\\ 2D LiDAR and radar measurements\end{tabular}  & 5,964              & Beam prediction                                                                              \\ \cline{2-6} 
			& 10-12              & Pedestrian                      & RGB images, GPS,  mmWave receive power                                                                             & 1,528              & \begin{tabular}[c]{@{}c@{}}Beam and blockage prediction, \\ user identification, positioning\end{tabular}                    \\ \hline
			2                  & 16                 & Fixed Wireless                  & RGB images, GPS,  mmWave receive power                                                                             & 9,827              & Blockage identification and prediction                                                                          \\ \hline
			\multirow{3}{*}{3} & 17-22              & \multirow{3}{*}{Fixed Wireless} & RGB images, GPS,  mmWave receive power                                                                             & 460,838           & Blockage identification and prediction                                                                          \\ \cline{2-2} \cline{4-6} 
			& 24-29              &                                 & \begin{tabular}[c]{@{}c@{}}RGB images, GPS,  mmWave receive power,\\ 2D LiDAR\end{tabular}                         & 500,000           & \begin{tabular}[c]{@{}c@{}} Blockage identification and prediction, \\ object detection/classification \end{tabular}                                                                        \\ \cline{2-2} \cline{4-6} 
			& 30                 &                                 & \begin{tabular}[c]{@{}c@{}}RGB images, GPS,  mmWave receive power,\\ 2D LiDAR, and radar measurements\end{tabular} & 14,660            & \begin{tabular}[c]{@{}c@{}} Blockage identification and prediction, \\ object detection/classification  \end{tabular}                                                                              \\ \hline
			4                  & 23                 & Drone Comm.                     & RGB images, GPS,  mmWave receive power                                                                             & 11,387            & Beam prediction                                                                              \\ \hline
			\multirow{2}{*}{5} & 31-34              & \multirow{2}{*}{V2I}            & \begin{tabular}[c]{@{}c@{}}RGB images, GPS,  mmWave receive power,\\ 3D LiDAR, and radar measurements\end{tabular} & 18,667            & \begin{tabular}[c]{@{}c@{}}Beam prediction, \\ user and blockage identification\end{tabular} \\ \cline{2-2} \cline{4-6} 
			& 35                 &                                 & \begin{tabular}[c]{@{}c@{}}RGB images, GPS,  mmWave receive power,\\ and radar measurements\end{tabular}           & 3,045              & \begin{tabular}[c]{@{}c@{}}Beam prediction, \\ user and blockage identification\end{tabular} \\ \hline
			6                  & 36-39              & V2V                             & \begin{tabular}[c]{@{}c@{}}RGB images, GPS,  mmWave receive power,\\ 3D LiDAR, and radar measurements\end{tabular} & 30,000           & \begin{tabular}[c]{@{}c@{}}Beam prediction, \\ user and blockage identification\end{tabular} \\ \hline
			7                  & 40                 & RIS                             & \begin{tabular}[c]{@{}c@{}}RGB images, GPS,  mmWave receive power,\\ 3D LiDAR, and radar measurements\end{tabular} & 5,000             & \begin{tabular}[c]{@{}c@{}}Beam prediction, \\ user and blockage identification\end{tabular} \\ \hline
		\end{tabular}
		\label{tab:data_statistics}
	\end{table*}
	

	DeepSense 6G dataset consists of real-world multi-modal datasets from collected multiple locations. As described in Section~\ref{sec:structure}, the data is provided as different scenarios where each scenario is one long data collection. It consists of $40$+ scenarios (and continuously growing) with realistic and challenging datasets to encourage the development of novel applications. Further, the scenarios are collected to enable varied deployments with sufficient variance. In this section, we highlight the inherent diversity of these scenarios and emphasize the value they add to the development of sensing-aided wireless communication applications.
	
	\textbf{Multiple sensors:} Each sensing modality has its advantages and limitations. For example, radar data is primarily suitable for uncrowded scenarios. Similarly, common positioning sensors generally do not provide accurate positions. The publicly available GPS data has an average error of $1-5$m. Utilizing multiple sensing modalities, i.e., data fusion, is a promising way to overcome such limitations. It is possible to develop more robust solutions by simultaneously leveraging multiple data modalities. However, developing such robust solutions requires the availability of large-scale multi-modal datasets. DeepSense 6G dataset takes the first step towards enabling the development of such novel solutions. It consists of $40$+ publicly available real-world scenarios with co-existing communication and multi-modal sensing data collected with a set of sensors such as RGB camera, 2D/3D LiDAR, radar, GPS RTK kit, or a subset of them.

	\textbf{Multiple locations:}  Data collection location plays a critical role in determining the quality of wireless and sensing data. In particular, both stationary and dynamic objects in the environment impact the performance of the $5$G and beyond wireless communication systems. Therefore, it is essential to select locations that can help in collecting meaningful and diverse data. The 40+ scenarios in DeepSense 6G dataset were collected in two countries (USA and Spain) with more than $15$+ indoor and outdoor locations. Each location was carefully selected to capture challenging and realistic scenarios. The outdoor locations have diverse characteristics; from quiet parking lots to busy downtown streets, the scenarios feature varying densities of pedestrians and vehicles, their mobility speeds, and environment geometry. Similarly, for the indoor locations, we increase diversity by selecting locations with different shapes and sizes, such as conference rooms, corridors, etc.

	\textbf{Different times of the day:} 
	Similar to the locations, the time of data collection plays an important role in determining the overall quality of the collected data. For example, cameras are susceptible to lighting conditions resulting in noisy images in low-light/dark scenarios. Such degradation in image quality can significantly impact the downstream tasks. Developing robust sensing-aided solutions that can perform efficiently in different lighting conditions requires the availability of large-scale real-world diverse datasets. For that, we collected data at different times of the day to incorporate such diversity in the DeepSense 6G dataset. In particular, scenarios $2, 4, 14, 17-19, 33$, and $34$ are all collected during the night, with plans to add more scenarios shortly. The remaining scenarios were all collected during the day but at different times to increase variance. With data collected from early morning to late night, the scenarios of DeepSense 6G cover the entire spectrum.

	\begin{figure*}[!t]
		\centering
		\includegraphics[width=0.75\linewidth]{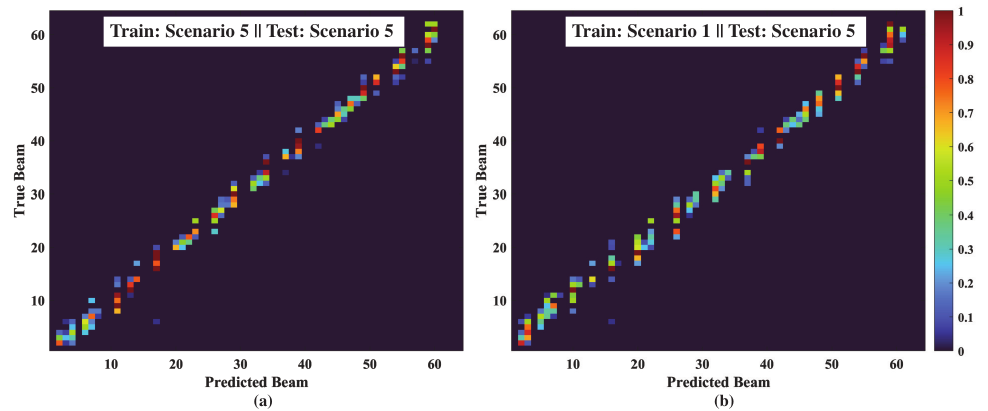}
		\caption{This figure plots the confusion matrices for the top-1 predicted beam indices. It shows the following two cases: (a) Training and testing on scenario $5$ development dataset and (b) training on scenario $1$ data and evaluating on scenario $5$ test dataset. The figure on the right, (b), shows that even without no training data of scenario $5$, the proposed solution can predict the optimal beam indices with sufficient accuracy.  }
		\label{fig:beam_pred_cm}
	\end{figure*}

	\textbf{Different weather conditions:} The performance of the different sensing modalities is affected by the weather conditions. For example, LiDAR data is severely impacted by poor weather conditions such as fog, dust, rain, and snow. Similarly, the visibility of cameras significantly reduces in foggy or rainy weather. Such challenges raise questions about the robustness of the solution developed based on these modalities. Developing efficient and reliable sensing-aided solutions and evaluating them requires data during different weather conditions. For this, we collect data in sunny, cloudy, windy, and rainy weather conditions and at various temperature and humidity levels. With our collaborators in other places, we plan to continue adding scenarios with diverse weather conditions. 
	
	\textbf{Different types of UEs:} Variance in object classes, object instances, and object speeds are crucial features of a dataset. Object classes and speeds are expected to significantly impact several downstream tasks, such as sensing-aided beam prediction, blockage prediction, and handoff. Developing solutions that can adapt and generalize across different deployment sites require dataset consisting of different real-world object classes (types). To that end, the different scenarios in DeepSense 6G dataset are designed to capture this inherent variability of real-world locations. To provide UE variability in the dataset, the scenarios capture people on their skateboards, bike, or scooters, traveling along the sidewalk, and crossing the street. Also, certain scenarios were collected on some of the busiest downtown streets, which helped capture various types of vehicles, such as sedans, SUVs, trucks, buses, etc., traveling at different speeds. All that diversity in traffic makes the dataset diverse from both sensing and wireless perspectives.
	
	\textbf{Deployment scenarios:} The DeepSense 6G dataset covers a wide range of deployment scenarios. In Fig~\ref{fig:scenarios}, we show some of the adopted deployment scenarios and the associated testbeds. In particular, as shown in this figure, the DeepSense 6G dataset includes the following deployment scenarios: Vehicle to infrastructure (V2I),  drone communication, vehicle to vehicle (V2V), reconfigurable intelligent surfaces (RIS), indoor use cases, and pedestrians, to name a few. The sheer range of the deployment scenarios that the DeepSense 6G dataset incorporates highlights its potential for enabling various critical use cases and applications. 
	
	
	\section{Enabled Applications and ML Tasks} \label{sec:deepsense_applications}
			
	The DeepSense 6G dataset, with its diverse scenarios, enables a wide range of applications in the interplay of communications, sensing, and positioning. This includes, as examples, multi-modal sensing aided beam and channel prediction, proactive blockage prediction and hand-off, waveform design for ISAC systems, radar/LiDAR object detection and classification, and mmWave-based positioning. To accelerate the machine learning research in these directions and to facilitate results reproducibility, benchmarking, and comparisons, we also provide, as part of the DeepSense 6G dataset, task-specific development datasets for a number of key applications. It is important to note here that thanks to the unified approach in collecting and structuring the data of the different scenarios, the task-specific datasets can be constructed via combining data from multiple scenarios. This enables advanced machine learning research in transfer learning, generalizability, scalability, robustness, and distribution shift analysis, among others.

	\textbf{Example - Generalizability evaluation for ML-based beam prediction.} Now, we describe how to use the dataset in studying one important research problem, namely multi-modal vision and position-aided beam prediction. In particular, we study the generalizability of the developed beam prediction solution: Can a machine learning model be trained on a scenario dataset and then successfully generalize to other scenarios the model has not seen before? Therefore, given a multi-modal dataset consisting of communication and sensing data collected at different locations, the objective is to develop a machine learning-based solution that can adapt to unseen scenarios that are not part of the training dataset. We adopt the machine-learning model proposed in \cite{Charan2022} with a slight modification. Instead of utilizing a convolutional neural network (CNN)-based model for extracting meaningful features from the visual data, we utilize a state-of-the-art object detection model, YOLOv3 \cite{joseph2018a}, to detect and extract the bounding-box center coordinates of the different objects of interest in the visual data. The bounding-box coordinates and the normalized position data are then provided as input to a 2-layered feed-forward neural network to predict the optimal beam indices from a pre-defined beamforming codebook. 
	
	\textbf{Performance:} To evaluate the performance of the proposed solution, we adopt scenarios $1$ and $5$ of the DeepSense 6G dataset. These scenarios consist of diverse wireless, visual, and position data. They were collected at two locations and at different times of the day and in different lighting conditions, making them good candidates for studying this generalizability problem. The final development dataset consists of $2411$ and $2300$ samples from scenarios $1$ and $5$. It comprises pairs of visual and position data along with the ground-truth beam indices. The data from each scenario is further divided into training and validation sets following a split of $70-30\%$. Fig.~\ref{fig:beam_pred_cm} plots the confusion matrices for the following two cases: (a) Training and testing on scenario $5$ development dataset and (b) training on scenario 1 data and evaluating on scenario 5 test dataset. This figure shows that even with no training data from scenario 5, the proposed machine learning-based solution can still predict the optimal beams with sufficient accuracy (more than 70\% top-3 accuracy). Fig.~\ref{fig:beam_pred_cm} also shows that even when the model predicts a wrong beam index, this index is with high probability close to the optimal ground-truth beams and hence may still have good receive power. These results highlight the value of using the DeepSense 6G dataset to obtain practical insights about the performance of a developed machine learning solution in realistic communication environments.

	\section{Conclusion}
	This article presented the DeepSense 6G dataset, which is a large-scale real-world dataset comprising co-existing and synchronized multi-modal sensing and communication data from more than 40 deployment scenarios, including vehicle-to-infrastructure, vehicle-to-vehicle, reconfigurable intelligent surfaces, pedestrians, and drone communication. The article described the motivation behind the dataset, its scalable structure, the modular testbed definition, the available scenarios, and some of the machine learning applications enabled by the dataset. The goal of the DeepSense 6G dataset project is to advance the deep learning research and development in a wide range of applications in the interplay of communication and sensing to enable the reproduction and benchmarking of their datasets and research results.

	\balance


	\begin{IEEEbiographynophoto}
		{Ahmed Alkhateeb} received his Ph.D. degree in Electrical Engineering from The University of Texas at Austin, USA, in August 2016. Between Sept. 2016 and Dec. 2017, he was a Wireless Communications Researcher at the Connectivity Lab, Facebook. He joined Arizona State University (ASU) in spring 2018, where he is currently an Assistant Professor in the School of Electrical, Computer and Energy Engineering. His research interests are in the broad areas of wireless communications,  signal processing, machine learning, and applied math. Dr. Alkhateeb is the recipient of the 2016 IEEE Signal Processing Society Young Author Best Paper Award, and the 2021 NSF CAREER Award.
	\end{IEEEbiographynophoto}
	\vskip -2.5\baselineskip plus -1fil
	\begin{IEEEbiographynophoto}
		{Gouranga Charan} received his B.Tech. degree in Instrumentation Engineering from Indian Institute of Technology Kharagpur, India, in 2015 and M.S. degree in Electrical Engineering from Arizona State University, USA in 2021. He is currently pursuing the Ph.D. degree in Electrical Engineering with Arizona State University, Tempe, AZ, USA. His research interests include studying the different applications of deep learning in computer vision, wireless communications, and wireless sensing, with primary focus on sensing-aided wireless communication.
	\end{IEEEbiographynophoto}
	\vskip -2.5\baselineskip plus -1fil
	\begin{IEEEbiographynophoto}
		{Tawfik Osman} received the B.S. degree in Electrical and Electronics Engineering from Ashesi University, Berekuso, Ghana, in 2020 and M.S.E degree in Electrical Engineering from Arizona State University, USA in 2021. He is currently pursuing the Ph.D. degree in Electrical Engineering with Arizona State University. His research interests include the study and implementation of machine learning aided wireless communications systems and signal processing algorithms.
	\end{IEEEbiographynophoto} 
	\vskip -2.5\baselineskip plus -1fil
	\begin{IEEEbiographynophoto}
		{Andrew Hredzak} received the B.S degree in Electrical Engineering from Arizona State University, USA, in 2021. His research interests include the broad areas of wireless communications and signal processing.
	\end{IEEEbiographynophoto}
	\vskip -2.5\baselineskip plus -1fil
	\begin{IEEEbiographynophoto}
		{João Morais} received the B.Sc. and M.Sc. degrees in Electrical and Computer Engineering from University of Lisbon, Portugal, in 2018 and 2020. He is currently pursuing the Ph.D.  in Electrical Engineering with Arizona State University, USA. His research interests include signal processing for wireless communication networks, advanced antenna systems, and explainable AI.
	\end{IEEEbiographynophoto}
	\vskip -2.5\baselineskip plus -1fil
	\begin{IEEEbiographynophoto}
		{Umut Demirhan} received the B.S. and M.S. degrees in Electrical and Electronics Engineering from Bilkent University, Ankara, Turkey, in 2015 and 2017, respectively. He is currently pursuing the Ph.D.  in Electrical Engineering with Arizona State University, USA. His research interests include the broad areas of wireless communications, signal processing and machine learning.
	\end{IEEEbiographynophoto}
	\vskip -2.5\baselineskip plus -1fil
	\begin{IEEEbiographynophoto}
		{Nikhil Srinivas} received the B.Tech. in Electronics and Communications Engineering from SRM Institute of Science and Technology, Chennai, India in 2019. He is currently pursuing the M.S. degree in Electrical Engineering with Arizona State University, USA. His research interests include the different areas of wireless communications, estimation theory and machine learning.
	\end{IEEEbiographynophoto}

\end{document}